\documentclass{article}
\usepackage{amssymb}
\usepackage{amsmath}

\begin{document}

\title{Varying Constants}
\author{John D. Barrow \\
DAMTP\\
Centre for Mathematical Sciences\\
Cambridge University\\
Cambridge CB3 0WA\\
UK}
\maketitle

\begin{abstract}
We review properties of theories for the variation of the gravitation and
fine structure 'constants'. We highlight some general features of the
cosmological models that exist in these theories with reference to recent
quasar data that are consistent with time-variation in the fine structure
'constant' since a redshift of 3.5. The behaviour of a simple class of
varying-alpha cosmologies is outlined in the light of all the observational
constraints.
\end{abstract}

\section{ Introduction}

\bigskip\ \ 

There are several reasons why the possibility of varying constants should be
taken seriously \cite{con}. First, we know that the best candidates for
unification of the forces of nature in a quantum gravitational environment
only seem to exist in finite form if there are many more dimensions of space
than the three that we are familiar with. This means that the true constants
of nature are defined in higher dimensions and the three-dimensional
shadows\ we observe are no longer fundamental and do not need to be
constant. Any slow change in the scale of the extra dimensions would be
revealed by measurable changes in our three-dimensional 'constants'. Second,
we appreciate that some apparent constant might be determined partially or
completely by spontaneous symmetry-breaking processes in the very early
universe. This introduces an irreducibly random element into the values of
those constants. They may be different in different parts of the universe.
The most dramatic manifestation of this process is provided by the chaotic
and eternal inflationary universe scenarios where both the number and the
strength of forces in the universe at low energy can fall out differently in
different regions. Third, any outcome of a theory of quantum gravity will be
intrinsically probabilistic. It is often imagined that the probability
distributions for observables will be very sharply peaked but this may not
be the case for all possibilities. Thus, the value of the gravitation
'constant', $G,$ or its time derivative, $\dot{G},$ might be predicted to be
spatial random variables. Fourth, a non-uniqueness of the vacuum state for
the universe would allow other numerical combinations of the constants to
have occurred in different places. String theory indicates that there is a
huge 'landscape' ($>10^{500}$) of possible vacuum states that the universe
can find itself residing in as it expand and cools. Each will have different
constants and associated forces and symmetries. It is sobering to remember
that at present we have no idea why any of the constants of Nature take the
numerical values they do and we have never successfully predicted the value
of any dimensionless constant in advance of its measurement. Fifth, the
observational limits on possible variations are often very weak (although
they can be made to sound strong by judicious parametrisations). For
example, the cosmological limits on varying $G$ tell us only that $\dot{G}%
/G\leq 10^{-2}H_{0}$, where $H_{0}$ is the present Hubble rate. However, the
last reason to consider varying constants is currently the most compelling.
For the first time there is a body of detailed astronomical evidence for the
time variation of a traditional constant. The observational programme of
Webb et al \cite{webb1,webb2} has completed detailed analyses of three
separate quasar absorption line data sets taken at Keck and finds persistent
evidence consistent with the fine structure constant, $\alpha $, having been 
\textit{smaller} in the past, at $z=1-3.5.$ The shift in the value of $%
\alpha $ for all the data sets is given provisionally by $\Delta \alpha
/\alpha =(-0.57\pm $ $0.10)\times 10^{-5}.$ This result is currently the
subject of detailed analysis and reanalysis by the observers in order to
search for possible systematic biases in the astrophysical environment or in
the laboratory determinations of the spectral lines.

The first investigations of time-varying constants were those made by Lord
Kelvin and others interested in possible time-variation of the speed of
light at the end of the nineteenth century. In 1935 Milne devised a theory
of gravity, of a form that we would now term 'bimetric', in which there were
two times -- one ($t$) for atomic phenomena, one ($\tau $) for gravitational
phenomena -- linked by $\tau =\log (t/t_{0})$. Milne \cite{mil} required
that the 'mass of the universe' (what we would now call the mass inside the
particle horizon $M\ \approx c^{3}G^{-1}t$) be constant. This required $%
G\varpropto t.$ Interestingly, in 1937 the biologist J.B.S. Haldane took a
strong interest in this theory and wrote several papers \cite{hal} exploring
its consequences for the evolution of life. The argued that biochemical
activation energies might appear constant on the $t$ timescale yet increase
on the $\tau $ timescale, giving rise to a non-uniformity in the
evolutionary process. Also at this time there was widespread familiarity
with the mysterious 'large numbers' $O(10^{40})$ and $O(10^{80})$ through
the work of Eddington (although they had first been noticed by Weyl \cite%
{weyl} -- see ref. \cite{bt} and \cite{con} for the history). These two
ingredients were merged by Dirac in 1937 in a famous development (supposedly
written on his honeymoon) that proposed that these large numbers $(10^{40})$
were actually equal, up to small dimensionless factors. Thus, if we form $%
N\sim c^{3}t/Gm_{\text{n}}\sim 10^{80}$, the number of nucleons in the
visible universe, \ and equate it to the square of $N_{1}\sim e^{2}/Gm_{%
\text{n}}^{2}\sim 10^{40},$ the ratio of the electrostatic and gravitational
forces between two protons then we are led to conclude that one of the
constants, $e,G,c,h,m_{\text{n}}$ must vary with time. Dirac \cite{Dir}
chose $G\varpropto t^{-1}$ to carry the time variation. Unfortunately, this
hypothesis did not survive very long. Edward Teller \cite{tell} pointed out
that such a steep increase in $G$ to the past led to huge increases in the
Earth's surface temperature in the past. The luminosity of the sun varies as 
$L\varpropto G^{7}$ and the radius of the Earth's orbit as $R\varpropto
G^{-1}$ so the Earth's surface temperature $T_{\oplus }$ varies as $%
(L/R^{2})^{1/4}\varpropto G^{9/4}\varpropto t^{-9/4}$ and would exceed the
boiling point of water in the pre-Cambrian era. Life would be eliminated.
Gamow subsequently suggested that the time variation needed to reconcile the
large number coincidences be carried by $e$ rather than $G$, but again this
strong variation was soon shown to be in conflict with geophysical and
radioactive decay data. This chapter was brought to an end by Dicke \cite%
{dicke} who pointed out that the $N\sim N_{1}^{2}$ large number coincidence
was just the statement that $t$, the present age of the universe when our
observations are being made, is of order the main-sequence stellar lifetime, 
$t_{\text{ms}}\sim (Gm_{\text{n}}^{2}/hc)^{-1}h/m_{\text{n}}c^{2}\sim
10^{10} $ yrs, and therefore inevitable for observers made out of chemical
elements heavier than hydrogen and helium. Dirac never accepted this
anthropic explanation for the large number coincidences (believing that
'observers' would be present in the universe long after the stars had died)
but curiously can be found making exactly the same type of anthropic
argument to defend his own varying $G$ theory by highly improbable arguments
(that the Sun accretes material periodically during its orbit of the galaxy
and this extra material cancels out the effects of overheating in the past)
in correspondence with Gamow in 1967 (see \cite{con} for fuller details).

Dirac's proposal acted as a stimulus to theorists, like Jordan, Brans and
Dicke \cite{bd}, to develop rigorous theories which included the time
variation of $G$ self-consistently by modelling it as arising from the
space-time variation of some scalar field $\phi (\mathbf{x},t)$ whose motion
both conserved energy and momentum and created its own gravitational field
variations. In this respect the geometric structure of Einstein's equations
provides a highly constrained environment to introduce variations of
'constants'. Whereas in Newtonian gravity we are at liberty to introduce a
time-varying $G(t)$ into the law of gravity by

\begin{equation}
F=-\frac{G(t)Mm}{r^{2}}  \label{newt1}
\end{equation}

This creates a non-conservative dynamical system but can be solved fairly
straightforwardly \cite{jbnewt}. However, this strategy of simply 'writing
in' the variation of $G$ by merely replacing $G$ by $G(t)$ in the equations
that hold when $G$ is a constant fails in general relativity. If we were to
imagine the Einstein equations would generalise to ($G_{ab}$ is the Einstein
tensor)

\begin{equation}
G_{ab}=\frac{8\pi G(t)}{c^{4}}T_{ab}  \label{newt}
\end{equation}%
then taking a covariant divergence and using $\nabla ^{a}G_{ab}=0,$ together
with energy-momentum conservation ($\nabla ^{a}T_{ab}=0)$ requires that $%
\nabla G\equiv 0$ and no variations are possible in eq. (\ref{newt}).
Brans-Dicke theory is a familiar example of how the addition of an extra
piece to $T_{ab\text{ \ }}$together with the dynamics of a $G(\phi )$ fields
makes a varying $G$ theory possible. Despite the simplicity of this lesson
in the context of a varying $G$ theory it was not taken on board when
considering the variations of other non-gravitational constants and the
literature is full of limits on their possible variation which have been
derived by considering a theory in which the time-variation is just written
into the equations which hold when the constant does not vary. These
'limits' are clearly invalid but they will play an important role in guiding
us towards the areas where a full theory will find the strongest rigorous
bounds. Recently, the interest in the possibility that $\alpha $ varies in
time has led to the first extensive exploration of simple self-consistent
theories in which $a$ variations occur through the variation of some scalar
field.

\section{A Simple Varying-Alpha Theory}

\bigskip

We are going to consider some of the cosmological consequences of a simple
theory with time varying $\alpha .$ Such a theory was first formulated by
Bekenstein \cite{bek} as a generalisation of Maxwell's equations but
ignoring the consequences for the gravitational field equations. Recently,
Magueijo, Sandvik and myself have completed this theory \ \cite{bsbm,bsm1,
bsm2, bsm3, bsm4} to include the coupling to the gravitational sector and
analysed its general cosmological consequences. This theory considers only a
variation of the electromagnetic coupling and so far ignores any unification
with the strong and electroweak interactions. Extensions to include the weak
interaction via a generalised Weinberg-Salam theory have also been developed
recently, see refs. \cite{km, sb}.

Our aim in studying this theory is to build up understanding of the effects
of the expansion on varying $\alpha $ and to identify features that might
carry over into more general theories in which all the unified interactions
vary \cite{banks, guts, marc}. The constraint imposed on varying $\alpha $
by the need to bring about unification at high energy is likely to be
significant but the complexities of analysing the simultaneous variation of
all the constants involved in the supersymmetric version of the standard
model are considerable. At the most basic level we recognise that any time
variation in the fine structure could be carried by either or both of the
electromagnetic or weak couplings above the electroweak scale.

The idea that the charge on the electron, or the fine structure constant,
might vary in cosmological time was proposed in 1948 by Teller, \cite{tell},
who suggested that $\alpha \propto (\ln t)^{-1}$ was implied by Dirac's
proposal that $G\propto t^{-1}$ and the numerical coincidence that $\alpha
^{-1}\sim \ln (hc/Gm_{\text{p}})$, where $m_{\text{p }}$is the proton mass.
Later, in 1967, Gamow \cite{gam} suggested $\alpha \propto t$ as an
alternative to Dirac's time-variation of the gravitation constant, $G$, as a
solution of the large numbers coincidences problem and in 1963 Stanyukovich
had also considered varying $\alpha $, \cite{stan}, in this context.
However, this power-law variation in the recent geological past was soon
ruled out by other evidence \cite{dyson}.

There are a number of possible theories allowing for the variation of the
fine structure constant, $\alpha $. In the simplest cases one takes $c$ and $%
\hbar $ to be constants and attributes variations in $\alpha $ to changes in 
$e$ or the permittivity of free space (see \cite{am} for a discussion of the
meaning of this choice). This is done by letting $e$ take on the value of a
real scalar field which varies in space and time (for more complicated
cases, resorting to complex fields undergoing spontaneous symmetry breaking,
see the case of fast tracks discussed in \cite{covvsl}). Thus $%
e_0\rightarrow e=e_0\epsilon (x^\mu ),$ where $\epsilon $ is a dimensionless
scalar field and $e_0$ is a constant denoting the present value of $e$. This
operation implies that some well established assumptions, like charge
conservation, must give way \cite{land}. Nevertheless, the principles of
local gauge invariance and causality are maintained, as is the scale
invariance of the $\epsilon $ field (under a suitable choice of dynamics).
In addition there is no conflict with local Lorentz invariance or covariance.

With this set up in mind, the dynamics of our theory is then constructed as
follows. Since $e$ is the electromagnetic coupling, the $\epsilon $ field
couples to the gauge field as $\epsilon A_{\mu }$ in the Lagrangian and the
gauge transformation which leaves the action invariant is $\epsilon A_{\mu
}\rightarrow \epsilon A_{\mu }+\chi _{,\mu },$ rather than the usual $A_{\mu
}\rightarrow A_{\mu }+\chi _{,\mu }.$ The gauge-invariant electromagnetic
field tensor is therefore 
\begin{equation}
F_{\mu \nu }=\frac{1}{\epsilon }\left( (\epsilon A_{\nu })_{,\mu }-(\epsilon
A_{\mu })_{,\nu }\right) ,
\end{equation}%
which reduces to the usual form when $\epsilon $ is constant. The
electromagnetic part of the action is still 
\begin{equation}
S_{\text{em}}=-\int d^{4}x\sqrt{-g}F^{\mu \nu }F_{\mu \nu }.
\end{equation}%
and the dynamics of the $\epsilon $ field are controlled by the kinetic term 
\begin{equation}
S_{\epsilon }=-\frac{1}{2}\frac{\hslash }{l^{2}}\int d^{4}x\sqrt{-g}\frac{%
\epsilon _{,\mu }\epsilon ^{,\mu }}{\epsilon ^{2}},
\end{equation}%
as in dilaton theories. Here, $l$ is the characteristic length scale of the
theory, introduced for dimensional reasons. This constant length scale gives
the scale down to which the electric field around a point charge is
accurately Coulombic. The corresponding energy scale, $\hbar c/l,$ has to
lie between a few tens of MeV and Planck scale, $\sim 10^{19}$GeV to avoid
conflict with experiment.

Our generalisation of the scalar theory proposed by Bekenstein \cite{bek}
described in refs. \cite{bsm1, bsm2, bsm3, bsm4} includes the gravitational
effects of $\psi $ and gives the field equations: 
\begin{equation}
G_{\mu \nu }=8\pi G\left( T_{\mu \nu }^{\text{matter}}+T_{\mu \nu }^{\psi
}+T_{\mu \nu }^{\text{em}}e^{-2\psi }\right) .  \label{ein}
\end{equation}%
The stress tensor of the $\psi $ field is derived from the lagrangian $%
\mathcal{L}_{\psi }=-{\frac{\omega }{2}}\partial _{\mu }\psi \partial ^{\mu
}\psi $ and the $\psi $ field obeys the equation of motion 
\begin{equation}
\square \psi =\frac{2}{\omega }e^{-2\psi }\mathcal{L}_{\text{em}}
\label{boxpsi}
\end{equation}%
where we have defined the coupling constant $\omega =(c)/l^{2}$. This
constant is of order $\sim 1$ if, as in \cite{bsbm}, the energy scale is
similar to Planck scale. It is clear that $\mathcal{L}_{\text{em}}$ vanishes
for a sea of pure radiation since then $\mathcal{L}_{\text{em}%
}=(E^{2}-B^{2})/2=0$. We therefore expect the variation in $\alpha $ to be
driven by electrostatic and magnetostatic energy-components rather than
electromagnetic radiation.

In order to make quantitative predictions we need to know how much of the
non-relativistic matter contributes to the RHS of Eqn.~(\ref{boxpsi}). This
is parametrised by $\zeta \equiv \mathcal{L}_{\text{em}}/\rho $, where $\rho 
$ is the energy density, and for baryonic matter $\mathcal{L}_{\text{em}%
}=E^{2}/2$. For protons and neutrons $\zeta _{\text{p}}$ and $\zeta _{\text{n%
}}$ can be \textit{estimated} from the electromagnetic corrections to the
nucleon mass, $0.63$ MeV and $-0.13$ MeV, respectively \cite{zal}. This
correction contains the $E^{2}/2$ contribution (always positive), but also
terms of the form $j_{\mu }a^{\mu }$ (where $j_{\mu }$ is the quarks'
current) and so cannot be used directly. Hence we take a guiding value $%
\zeta _{\text{p}}\approx \zeta _{\text{n}}\sim 10^{-4}$. Furthermore the
cosmological value of $\zeta $ (denoted $\zeta _{\text{m}}$) has to be
weighted by the fraction of matter that is non-baryonic. Hence, $\zeta _{%
\text{m}}$ depends strongly on the nature of the dark matter and can take
both positive and negative values depending on which of Coulomb-energy or
magnetostatic energy dominates the dark matter of the Universe. It could be
that $\zeta _{\text{CDM}}\approx -1$ (superconducting cosmic strings, for
which $\mathcal{L}_{\text{em}}\approx -B^{2}/2)$, or $\zeta _{\text{CDM}}\ll
1$ (neutrinos). BBN predicts an approximate value for the baryon density of $%
\Omega _{\text{B}}\approx 0.03$ (where $\Omega _{\text{B}}$ is the density
of matter in units of the critical density $3H^{2}/8\pi G$) with a Hubble
parameter of $\ H=60$ Kms$^{\text{-1}}$ Mpc$^{\text{-1}}$, implying $\Omega
_{\text{CDM}}\approx 0.3$. Thus depending on the nature of the dark matter $%
\zeta _{\text{m}}$ can be virtually anything between $-1$ and $+1$. The
uncertainties in the underlying quark physics and especially the
constituents of the dark matter make it difficult to impose more certain
bounds on $\zeta _{\text{m}}$.

We should not confuse this theory with other similar variations.
Bekenstein's theory does not take into account the stress energy tensor of
the dielectric field in Einstein's equations. Dilaton theories predict a
global coupling between the scalar and all other matter fields (not just the
electromagnetically charged material) \cite{cham, bass, mart, beknew, dm}.
As a result they predict variations in other constants of nature, and also a
different cosmological dynamics.

\subsection{The cosmological equations}

Assuming a homogeneous and isotropic Friedmann metric with expansion scale
factor $a(t)$ and curvature parameter $k$ in eqn. (\ref{ein}), we obtain the
field equations ($c\equiv 1$) 
\begin{eqnarray}
\left( \frac{\dot{a}}{a}\right) ^{2} &=&\frac{8\pi G}{3}\left( \rho _{\text{m%
}}\left( 1+\zeta _{\text{m}}\exp {[-2\psi ]}\right) +\rho _{\text{r}}\exp {%
[-2\psi ]}+\frac{\omega }{2}\dot{\psi}^{2}\right)  \nonumber \\
&&-\frac{k}{a^{2}}+\frac{\Lambda }{3},  \label{fried1}
\end{eqnarray}%
where $\Lambda $ is the cosmological constant. For the scalar field we have
the propagation equation, 
\begin{equation}
\ddot{\psi}+3H\dot{\psi}=-\frac{2}{\omega }\exp {[-2\psi ]}\zeta _{_{\text{m}%
}}\rho _{\text{m}},  \label{psidot}
\end{equation}%
where $H\equiv \dot{a}/a$ is the Hubble expansion rate$.$ We can rewrite
this more simply as

\begin{equation}
(\dot{\psi}a^{3}\dot{)}=N\exp [-2\psi ]  \label{psidot2}
\end{equation}%
where $N$ is a positive constant defined by

\begin{equation}
N=-\frac{2\zeta _{\text{m}}\rho _{\text{m}}a^{3}}{\omega }  \label{N}
\end{equation}

Note that the sign of the evolution of $\psi $ is dependent on the sign of $%
\zeta _{\text{m}}$. Since the observational data is consistent with a \emph{%
smaller} value of $\alpha $ in the past, we will in this paper confine our
study to \emph{negative} values of $\zeta _{\text{m}}$, in line with our
recent discussion in Refs. \cite{bsbm, bsm1, bsm2, bsm3, bsm4}. The
conservation equations for the non-interacting radiation and matter
densities are 
\begin{eqnarray}
\dot{\rho _{\text{m}}}+3H\rho _{\text{m}} &=&0 \\
\dot{\rho _{\text{r}}}+4H\rho _{r} &=&2\dot{\psi}\rho _{r}.
\label{conservation}
\end{eqnarray}%
and so $\rho _{\text{m}}\propto a^{-3}$ and $\rho _{\text{r}}$ $e^{-2\psi
}\propto a^{-4},$ respectively. If additional non-interacting perfect fluids
satisfying equation of state $p=(\gamma -1)\rho $ are added to the universe
then they contribute density terms $\rho \propto a^{-3\gamma }$ to the RHS
of eq.(\ref{fried1}) as usual. This theory enables the cosmological
consequences of varying $e$, to be analysed self-consistently rather than by
changing the constant value of $e$ in the standard theory to another
constant value, as in the original proposals made in response to the large
numbers coincidences.

We have been unable to solve these equations in general except for a few
special cases. However, as with the Friedmann equation of general
relativity, it is possible to determine the overall pattern of cosmological
evolution in the presence of matter, radiation, curvature, and positive
cosmological constant by matched approximations. We shall consider the form
of the solutions to these equations when the universe is successively
dominated by the kinetic energy of the scalar field $\psi $, pressure-free
matter, radiation, negative spatial curvature, and positive cosmological
constant$.$ Our analytic expressions are checked by numerical solutions of (%
\ref{fried1}) and (\ref{psidot}).

There are a number of conclusions that can be drawn from the study of the
simple BSBM models with $\zeta _{\text{m}}<0$. These models give a good fit
to the varying $\alpha $ implied by the QSO data of refs. \cite{webb1,webb2}%
. There is just a single parameter to fit and this is given by the choice

\begin{equation}
-\frac{\zeta _{\text{m}}}{\omega }=(2\pm 1)\times 10^{-4}  \label{om}
\end{equation}

The simple solutions predict a slow (logarithmic) time increase during the
dust era of $k=0$ Friedmann universes. The cosmological constant turns off
the time-variation of $\alpha $ at the redshift when the universe begins to
accelerate ($z\sim 0.7$) and so there is no conflict between the $\alpha $
variation seen in quasars at $z\sim 1-3.5$ and the limits on possible
variation of $\alpha $ deduced from the operation of the Oklo natural
reactor \cite{oklo} (even assuming that the cosmological variation applies
unchanged to the terrestrial environment). The reactor operated 1.8 billion
years ago at a redshift of only $z\sim 0.1$ when no significant variations
were occurring in $\alpha $. The slow logarithmic increase in $\alpha $ also
means that we would not expect to have seen any effect yet in the anisotropy
of the microwave backgrounds \cite{bat, avelino}: the value of $\alpha $ at
the last \ scattering redshift, $z=1000,$ is only 0.005\% lower than its
value today. Similarly, the essentially constant evolution of $\alpha $
predicted during the radiation era leads us to expect no measurable effects
on the products of Big Bang nucleosynthesis (BBN) \cite{jdb} because $\alpha 
$ was only 0.007\% smaller at BBN than it is today. This does not rule out
the possibility that unification effects in a more general theory might
require variations in weak and strong couplings, or their contributions to
the neutron-proton mass difference, which might produce observable
differences in the light element productions and new constraints on varying $%
\alpha $ at $z\sim 10^{9}-10^{10}.$ By contrast, varying-alpha cosmologies
with $\zeta >0$ lead to bad consequences unless the scalar field driving the
alpha variations is a 'ghost' field, with negatively coupled kinetic energy,
in which case there are interesting cosmological consequences, \cite{bkm}.
The fine structure falls rapidly at late times and the variation is such
that it even comes to dominate the Friedmann equation for the cosmological
dynamics. We regard this as a signal that such models are astrophysically
ruled out and perhaps also mathematically badly behaved.

We should also mention that theories in which $\alpha $ varies will in
general lead to violations of the weak equivalence principle (WEP). This is
because the $\alpha $ variation is carried by a field like $\psi $ and this
couples differently to different nuclei because they contain different
numbers of electrically charged particles (protons). The theory discussed
here has the interesting consequence of leading to a relative acceleration
of order $10^{-13}$ \cite{bmswep} if the free coupling parameter is fixed to
the value given in eq. (\ref{om}) using a best fit of the theories
cosmological model to the QSO observations of refs. \cite{webb1, webb2}.
Other predictions of such WEP violations have also been made in refs. \cite%
{poly, zal, zald, dam}. The observational upper bound on this parameter from
direct experiment is just an order of magnitude larger, at $10^{-12},$ and
limits from the motion of the Moon are of similar order, but space-based
tests planned for the STEP mission are expected to achieve a sensitivity of
order $10^{-18}$ and will provide a completely independent check on theories
of time-varying $e$ and $\alpha .$This is an exciting prospect for the
future.

\subsection{The nature of the Friedmann solutions}

Let us present the predicted cosmological evolution of $\alpha $ in the BSBM
theory$,$ that we summarised above, in a little more detail. During the
radiation era the expansion scale factor of the universe increases as $%
a(t)\sim t^{1/2}$ and $\alpha $ is essentially constant in universes with an
entropy per baryon and present value of $\alpha $ like our own. It increases
in the dust era, where $a(t)\sim t^{2/3}$. The increase in $\alpha $
however, is very slow with a late-time solution for $\psi $ proportional to $%
\frac{1}{2}\log (2N\log (t))$, and so

\begin{equation}
\alpha \sim 2N\log t
\end{equation}

This slow increase continues until the expansion becomes dominated by
negative curvature, $a(t)\sim t$, or by a cosmological vacuum energy, $%
a(t)\sim \exp [\Lambda t/3]$. Thereafter $\alpha $ asymptotes rapidly to a
constant. If we set the cosmological constant equal to zero and $k=0$ then,
during the dust era, $\alpha $ would continue to increase indefinitely. $\ $%
The effect of the expansion is very significant at all times. If we were to
turn it off and set $a(t)$ constant then we could solve the $\psi $ equation
to give the following exponentially growing evolution for $\alpha $, \cite%
{bmota}$:$

\begin{equation}
\alpha =\exp [2\psi ]=A^{-2}\cosh ^{2}[AN^{1/2}(t+t_{0})];\text{ }A\text{
constant.}
\end{equation}

From these results it is evident that non-zero curvature or cosmological
constant brings to an end the increase in the value of $\alpha $ that occurs
during the dust-dominated era. Hence, if the spatial curvature and $\Lambda $
are both too\textit{\ small} it is possible for the fine structure constant
to grow too large for biologically important atoms and nuclei to exist in
the universe. There will be a time in the future when $\alpha $ reaches too
large a value for life to emerge or persist. The closer a universe is to
flatness or the closer $\Lambda $ is to zero so the longer the monotonic
increase in $\alpha $ will continue, and the more likely it becomes that
life will be extinguished. Conversely, a non-zero positive $\Lambda $ or a
non-zero negative curvature will stop the increase of $\alpha $ earlier and
allow life to persist for longer. If life can survive into the curvature or $%
\Lambda $-dominated phases of the universe's history then it will \ \ \ \ \
\ \ \ not be threatened by the steady cosmological increase in $\alpha $
unless the universe collapses back to high density.

There have been several studies, following Carter, \cite{car} and Tryon \cite%
{try}, of the need for life-supporting universes to expand close to the
'flat' Einstein de Sitter trajectory for long periods of time. This ensures
that the universe cannot collapse back to high density before galaxies,
stars, and biochemical elements can form by gravitational instability, or
expand too fast for stars and galaxies to form by gravitational instability 
\cite{ch, bt}. Likewise, it was pointed out by Barrow and Tipler, \cite{bt}
that there are similar anthropic restrictions on the magnitude of any
cosmological constant, $\Lambda $. If it is too large in magnitude it will
either precipitate premature collapse back to high density (if $\Lambda <0$)
or prevent the gravitational condensation of any stars and galaxies (if $%
\Lambda >0$). Thus, we can provide good anthropic reasons why we can expect
to live in an old universe that is neither too far from flatness nor
dominated by a much stronger cosmological constant than observed ($%
\left\vert \Lambda \right\vert \leq 10\left\vert \Lambda _{\text{obs}%
}\right\vert $).

Inflationary universe models provide a possible theoretical explanation for
proximity to flatness but no explanation for the smallness of the
cosmological constant. Varying speed of light theories \cite{moffat, am, ba,
bm, bhvsl} offer possible explanations for proximity to flatness and
smallness of a classical cosmological constant (but not necessarily for one
induced by vacuum corrections in the early universe). We have shown that if
we enlarge our cosmological theory to accommodate variations in some
traditional constants then\textit{\ }it appears to be anthropically
disadvantageous for a universe to lie too close to flatness or for the
cosmological constant to be too close to zero. This conclusion arises
because of the coupling between time-variations in constants like $\alpha $
and the curvature or $\Lambda $, which control the expansion of the
universe. The onset of a period of $\Lambda $ or curvature domination has
the property of dynamically stabilising the constants, thereby creating
favourable conditions for the emergence of structures. This point has been
missed in previous studies because they have never combined the issues of $%
\Lambda $ and flatness and the issue of the values of constants. By coupling
these two types of anthropic considerations we find that too small a value
of $\Lambda $ or the spatial curvature can be as poisonous for life as too
much. Universes like those described above, with increasing $\alpha (t),$
lead inexorably to an epoch where $\alpha $ is too large for the existence
of atoms, molecules, and stars to be possible \cite{bsm2}.

Surprisingly, there has been almost no consideration of habitability in
cosmologies with time-varying constants since Haldane's discussions \cite%
{hal} of the biological consequences of Milne's bimetric theory of gravity.
Since then, attention has focussed upon the consequences of universes in
which the constants are different but still constant. Those cosmologies with
varying constants that have been studied have not considered the effects of
curvature or $\Lambda $ domination on the variation of constants and have
generally considered power-law variation to hold for all times. The examples
described here show that this restriction has prevented a full appreciation
of the coupling between the expansion dynamics of the universe and the
values of the constants that define the course of local physical processes
within it. Our discussion of a theory with varying $\alpha $ shows for the
first time a possible reason why the 3-curvature of universes and the value
of any cosmological constant may need to be bounded \textit{below} in order
that the universe permit atomic life to exist for a significant period.
Previous anthropic arguments \cite{bt} have shown that the spatial curvature
of the universe and the value of the cosmological constant must be bounded 
\textit{above} in order for life-supporting environments (stars) to develop.
We note that the lower bounds discussed here are more fundamental than these
upper bounds because they derive from changes in $\alpha $ which have direct
consequences for biochemistry whereas the upper bounds just constrain the
formation of astrophysical environments by gravitational instability. Taken
together, these arguments suggest that within an ensemble of all possible
worlds where $\alpha $ and $G$ are time variables, there might only be a
finite interval of \textit{non-zero }values of the curvature and
cosmological constant contributions to the dynamics that both allow galaxies
and stars to form and their biochemical products to persist.

\section{The Observational Evidence\protect\bigskip}

\bigskip New precision studies of relativistic fine structure in the
absorption lines of dust clouds around quasars by Webb et al., \cite%
{webb1,webb2}, have led to widespread theoretical interest in the question
of whether the fine structure constant, $\alpha _{\text{em}}=e^{2}/\hbar c$,
has varied in time and, if so, how to accommodate such a variation by a
minimal perturbation of existing theories of electromagnetism.\ These
astronomical studies have proved to be more sensitive than laboratory probes
of the constancy of the fine structure 'constant', which currently give
bounds on the time variation of $\dot{\alpha}_{\text{em}}/\alpha _{\text{em}%
}\equiv -0.4\pm 16\times 10^{-16}$ yr$^{-1}$, \cite{lab1}, $\left\vert \dot{%
\alpha}_{\text{em}}/\alpha _{\text{em}}\right\vert \ <1.2\times 10^{-15}$ yr$%
^{-1}$, \cite{lab2}, $\dot{\alpha}_{\text{em}}/\alpha _{\text{em}}\equiv
-0.9\pm 2.9\times 10^{-16}$ yr$^{-1}$, \cite{lab3} by comparing atomic clock
standards based on different sensitive hyperfine transition frequencies, and 
$\dot{\alpha}_{\text{em}}/\alpha _{\text{em}}\equiv -0.3\pm 2.0\times
10^{-15}$ yr$^{-1}$ from comparing two standards derived from 1S-2S
transitions in atomic hydrogen after an interval of 2.8 years \cite{lab4}.
The quasar data analysed in refs. \cite{webb1,webb2} consists of three
separate samples of Keck-Hires observations which combine to give a data set
of 128 objects at redshifts $0.5<z<3$. The many-multiplet technique finds
that their absorption spectra are consistent with a shift in the value of
the fine structure constant between these redshifts and the present of $%
\Delta \alpha _{\text{em}}/\alpha _{\text{em}}\equiv \lbrack \alpha _{\text{%
em}}(z)-\alpha _{\text{em}}]/\alpha _{\text{em}}=-0.57\pm 0.10\times
10^{-5}, $ where $\alpha _{\text{em}}\equiv $ $\alpha _{\text{em}}(0)$ is
the present value of the fine structure constant \cite{webb1,webb2}.
Extensive analysis has yet to find a selection effect that can explain the
sense and magnitude of the relativistic line-shifts underpinning these
deductions. Further observational studies have been published in refs. \cite%
{chand1,chand2} using a different but smaller data set of 23 absorption
systems in front of 23 VLT-UVES quasars at $0.4\leq z\leq 2.3$ and have been
analysed using an approximate form of the many-multiplet analysis techniques
introduced in refs. \cite{webb1,webb2}. They obtained $\Delta \alpha _{\text{%
em}}/\alpha _{\text{em}}\equiv -0.6\pm 0.6\times 10^{-6}$; a figure that
disagrees with the results of refs. \cite{webb1,webb2} . However, reanalysis
is needed in order to understand the accuracy being claimed and ensure that
all spectral lines are being identified. Other observational studies of
lower sensitivity have also been made using OIII emission lines of galaxies
and quasars. The analysis of data sets of 42 and 165 quasars from the SDSS
gave the constraints $\Delta \alpha _{\text{em}}/\alpha _{\text{em}}\equiv
0.51\pm 1.26\times 10^{-4}$ and $\Delta \alpha _{\text{em}}/\alpha _{\text{em%
}}\equiv 1.2\pm 0.7\times 10^{-4}$ respectively for objects in the redshift
range $0.16\leq z\leq 0.8$ \cite{sdss}. Observations of a single quasar
absorption system at $z=1.15$ by Quast et al \cite{qu} gave $\Delta \alpha _{%
\text{em}}/\alpha _{\text{em}}\equiv -0.1\pm 1.7\times 10^{-6}$ , and
observations of an absorption system at $z=1.839$ by Levshakov et al \cite%
{lev} gave $\Delta \alpha _{\text{em}}/\alpha _{\text{em}}\equiv 2.4\pm
3.8\times 10^{-6}$. A preliminary analysis of constraints derived from the
study of the OH microwave transition from a quasar at $z=0.2467$, a method
proposed by Darling \cite{darl}, has given $\Delta \alpha _{\text{em}%
}/\alpha _{\text{em}}\equiv 0.51\pm 1.26\times 10^{-4}$, \cite{oh}$.$A
comparison of redshifts measured using molecules and atomic hydrogen in two
cloud systems by Drinkwater et al \cite{drink} at $z=0.25$ and $z=0.68$ gave
a bound of $\Delta \alpha _{\text{em}}/\alpha _{\text{em}}<5\times 10^{-6}$
and an upper bound on spatial variations of $\delta \alpha _{em}/\alpha _{%
\text{em}}<3\times 10^{-6}$ over 3 Gpc at these redshifts. A new study
comparing UV absorption redshifted into the optical with redshifted 21 cm
absorption lines from the same cloud in a sample of 8 quasars by Tzanavaris
et al \cite{tz}. This comparison probes the constancy of $\alpha ^{2}g_{%
\text{p}}m_{\text{e}}/m_{\text{p}}$ and gives $\Delta \alpha _{\text{em}%
}/\alpha _{\text{em}}\equiv 0.18\pm 0.55\times 10^{-5}$ if we assume that
the electron-proton mass ratio and proton $g$-factor, $g_{p}$, are both
constant.

Observational bounds derived from the microwave background radiation
structure \cite{cmb} and Big Bang nucleosynthesis \cite{jdb, bbn} are not
competitive at present (giving $\Delta \alpha _{\text{em}}/\alpha _{\text{em}%
}\lesssim 10^{-2}$ at best at $z\sim 10^{3}$ and $z\sim 10^{9}-10^{10}$)
with those derived from quasar studies, although they probe much higher
redshifts.

Other bounds on the possible variation of the fine structure constant have
been derived from geochemical studies, although they are subject to awkward
environmental uncertainties. The resonant capture cross-section for thermal
neutrons by samarium-149 about two billion years ago ($z\simeq 0.15$) in the
Oklo natural nuclear reactor has created a samarium-149:samarium-147 ratio
at the reactor site that is depleted by the capture process $^{149}$Sm$+$n $%
\rightarrow $ $^{150}$Sm$+\gamma $ to an observed value of only about 0.02
compared to the value of about 0.9 found in normal samples of samarium. The
need for this capture resonance to be in place two billion years ago at an
energy level within about $90$ meV of its current value leads to very strong
bounds on all interaction coupling constants that contribute to the energy
level, as first noticed by Shlyakhter \cite{shly, con}. The latest analyses
by Fujii et al \cite{fuj} allow two solutions (one consistent with no
variation the other with a variation) because of the double-valued form of
the capture cross-section's response to small changes in the resonance
energy over the range of possible reactor temperatures: $\Delta \alpha _{%
\text{em}}/\alpha _{\text{em}}\equiv -0.8\pm 1.0\times 10^{-8}$ or $\Delta
\alpha _{\text{em}}/\alpha _{\text{em}}\equiv 8.8\pm 0.7\times 10^{-8}.$ The
latter possibility does not include zero but might be excluded by further
studies of other reactor abundances. Subsequently, Lamoureax \cite{lam} has
argued that a better (non-Maxwellian) assumption about the thermal neutron
spectrum in the reactor leads to $6\sigma $ lower bound on the variation of $%
\Delta \alpha _{\text{em}}/\alpha _{\text{em}}>4.5\times 10^{-8}$ at $%
z\simeq 0.15$.

Studies of the effects of varying a fine structure constant on the $\beta $%
-decay lifetime was first considered by Peebles and Dicke \cite{PD} as a
means of constraining allowed variations in $\alpha _{\text{em}}$ by
studying the ratio of rhenium to osmium in meteorites. The $\beta $-decay \ $%
_{75}^{187}$Re$\rightarrow $ $_{76}^{187}$Os$+\bar{\nu}_{\text{e}}+$e$^{-}$
is very sensitive to $\alpha _{\text{em}}$ and the analysis of new
meteoritic data together with new laboratory measurements of the decay rates
of long-lived beta isotopes has led to a time-averaged limit of $\Delta
\alpha _{\text{em}}/\alpha _{\text{em}}=8\pm 16\times 10^{-7}$ \cite{Olive,
olivepos} for a sample that spans the age of the solar system ($z\leq 0.45$%
). Both the Oklo and meteoritic bounds are complicated by the possibility of
simultaneous variations of other constants which contribute to the energy
levels and decay rates; for reviews see refs. \cite{uzan, olive}. They also
apply to environments within virialised structures that do not take part in
the Hubble expansion of the universe and so it is not advisable to use them
in conjunction with astronomical information from quasars without a theory
that links the values of $\alpha _{\text{em}}$ in the two different
environments that differ in density by a factor of $O(10^{30}).$ Detailed
discussions of this problem when $G$ and $\alpha $ vary have been made in
refs. \cite{Barrow:2001, Mota:2004, Mota:2003}. \textit{\ }

\section{The Role of Inhomogeneities}

\bigskip All early studies of the cosmological consequences of varying
constants have assumed that they vary homogeneously. Such an assumption is
also implicit when laboratory data or solar system observations are used to
constrain cosmological theories of varying $G$ and $\alpha $. In reality
such a simple approach is very dangerous. Our local observations are made
inside a gross cosmological overdensity -- $10^{30}$ times denser than the
mean density of the background universe -- that is not taking part in the
universal expansion. We should no more expect laboratory observations of the
constancy of $\alpha $ to reflect what is happening on extragalactic scales
than we should expect a measurement of the density of the Earth to give a
good estimate of the density of the universe. In order to use our local
observations effectively we need a theoretical description of how variations
in, say, $\alpha $ will vary with the local density of matter as a result of
the process of galaxy, star, and planetary formation. For example, when a
cosmological overdensity separates out from the expansion of the universe,
and collapses under its own gravity, it will eventually come into a
stationary virial equilibrium. If $\alpha $ is a space-time variable it will
continue changing in the background universe after it has ceased to change
in the virialised protogalaxy with a density contrast of about $10^{6}$ with
respect to the background universe. In this way we see that the process of
galaxy formation leads us to expect that any time variation in fundamental
constants will be inevitably accompanied by a space variation that is
potentially much more marked. In particular both $\alpha $ and $\dot{\alpha}$
will exhibit different values inside and outside galaxies and galaxy
clusters. Moreover, we expect the residual time variations inside galaxies
(and hence in terrestrial laboratories) to be significantly smaller than
those to be found in extragalactic systems that take part in the expansion
of the universe \cite{Mota:2004, Mota:2003}. In contrast, if we go to very
large scales where we are observing very small fluctuations long before they
collapse into clusters and galaxies, we can calculate the effects of
spatially varying 'constants' \ on the isotropy of the microwave background
radiation. The author has recently shown \cite{spatial} that the evolution
eqn. (\ref{boxpsi}) means that spatial variations in $\alpha $ are driven by
spatial variations in the matter density which in turn produce spatial
variations in the gravitational potential. These potential variations create
temperature anisotropies in the microwave background on large angular
scales. The observational bound on these variations from the COBE and WMAP
satellites allow us to conclude that in these theories spatial variations in 
$\alpha $ are bounded above by $\delta \alpha /\alpha <2\times 10^{-9}$.
Very strong bounds can be also derived in this way on allowed spatial
variations in $G$ and $m_{\text{e}}/m_{\text{p}}$and in Bran-Dicke theory
and the new theory for varying $m_{\text{e}}/m_{\text{p}}$ recently devised
by Barrow and Magueijo \cite{mu}.

These theoretical developments, together with the appearance of new
observational probes of the constants of physics at high redshift, coupled
with recent rapid progress in direct laboratory probes of the stability of
atomic systems that depend sensitively on the value of the fine structure
constant here and now, promise to create an exciting new focal point in our
quest to understand the nature (as well as the number) of the fundamental
constants of Nature.

\bigskip

\textbf{Acknowledgements} I would like to thank my collaborators Jo\~{a}o
Magueijo, H\aa vard Sandvik, John Webb, Michael Murphy, Dagny Kimberly and
David Mota for discussions and for their essential contributions to the work
described here.

\end{document}